\documentclass{article}
\usepackage{graphicx}
\usepackage{amsmath}

\newtheorem{theorem}{Theorem}

\newtheorem{definition}[theorem]{Definition}

\newtheorem{lemma}[theorem]{Lemma}

\begin{document}

\title{Laplace Operator in Networks of Thin Fibers: Spectrum Near the Threshold.}
\author{S. Molchanov, B. Vainberg \thanks{
The authors were supported partially by the NSF grant DMS-0405927.} \and %
Dept. of Mathematics, University of North Carolina at Charlotte, \and %
Charlotte, NC 28223, USA}
\date{}
\maketitle

\begin{abstract}
Our talk at Lisbon SAMP conference was based mainly on our recent results on
small diameter asymptotics for solutions of the Helmgoltz equation in
networks of thin fibers. These results were published in \cite{MV1}. The
present paper contains a detailed review of \cite{MV1} under some
assumptions which make the results much more transparent. It also contains
several new theorems on the structure of the spectrum near the threshold.
small diameter asymptotics of the resolvent, and solutions of the evolution
equation.
\end{abstract}


\bigskip

\section{Introduction}

The paper concerns the asymptotic spectral analysis of the wave problems in
systems of wave guides when the thickness $\varepsilon $ of the wave guides
is vanishing. In the simplest case, the problem is described by the
stationary wave (Helmholtz) equation
\begin{equation}
-\varepsilon ^{2}\Delta u=\lambda u,\text{ \ \ \ }x\in \Omega _{\varepsilon
},\text{ \ \ \ }Bu=0\text{ \ \ on }\partial \Omega _{\varepsilon }\text{,}
\label{h0}
\end{equation}
in a domain $\Omega _{\varepsilon }\subset R^{d},$ $d\geq 2,$ with
infinitely smooth boundary (for simplicity) which has the following
structure: $\Omega _{\varepsilon }$ is a union of a finite number of
cylinders $C_{j,\varepsilon }$ (which we shall call channels)$,$ $1\leq
j\leq N,$ of lengths $l_{j}$ with the diameters of cross-sections of order $%
O\left( \varepsilon \right) $ and domains $J_{1,\varepsilon },\cdots
,J_{M,\varepsilon }$ (which we shall call junctions) connecting the channels
into a network. It is assumed that the junctions have diameters of the same
order $O(\varepsilon )$. The boundary condition has the form: $B=1$ (the
Dirichlet BC) or $B=\frac{\partial u}{\partial n}$ (the Neumann BC) or $%
B=\varepsilon \frac{\partial u}{\partial n}+\alpha (x),$ where $n$ is the
exterior normal and\ the function $\alpha >0$ is real valued and does not
depend on the longitudinal (parallel to the axis) coordinate on the boundary
of the channels. One also can impose one type of BC on the lateral boundary
of $\Omega _{\varepsilon }$ and another BC on free ends (which are not
adjacent to a junction) of the channels. For simplicity we assume that only
Dirichlet or Neumann BC are imposed on the free ends of the channels.
Sometimes we shall denote the operator $B$ on the lateral surface of $\Omega
_{\varepsilon }$ by $B_{0},$ and we shall denote the operator $B$ on the
free ends of the channels by $B_{e}.$

Let $m$ channels have infinite length. We start the numeration of $%
C_{j,\varepsilon }$ with the infinite channels. So, $l_{j}=\infty $ for $%
1\leq j\leq m.$ The axes of the channels form edges $\Gamma _{j},$ $1\leq
j\leq N,$ of the limiting $\left( \varepsilon \rightarrow 0\right) $ metric
graph $\Gamma $. We split the set $V$ of vertices $v_{j}$ of the graph in
two subsets $V=V_{1}\cup V_{2},$ where the vertices from the set $V_{1}$
have degree $1$ and the vertices from the set $V_{2}$ have degree at least
two, i.e. vertices $v_{i}\in V_{1}$ of the graph $\Gamma $ correspond to the
free ends of the channels, and vertices $v_{j}\in V_{2}$ correspond to the
junctions $J_{j,\varepsilon }$.

\begin{figure}[htbp]
\begin{center}
\includegraphics[width=0.8\columnwidth]{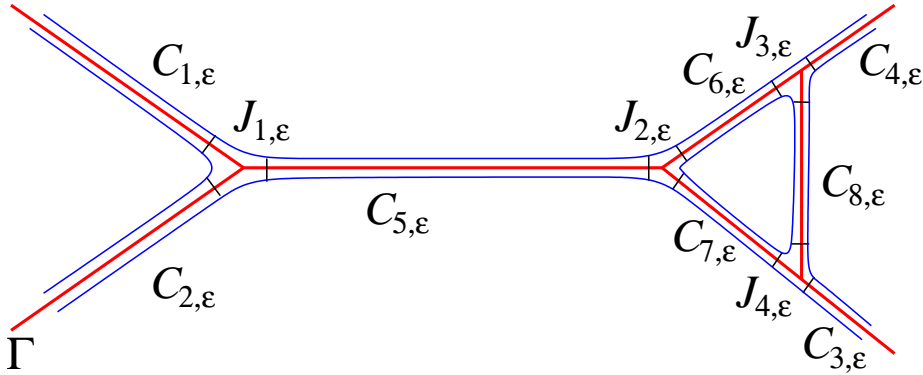}
\end{center}
\caption{An example of a domain $\Omega _{\protect\varepsilon }$ with four
junctions, four unbounded channels and four bounded channels.}
\label{fig-1}
\end{figure}

Equation (\ref{h0}) degenerates when $\varepsilon =0.$ One could omit $%
\varepsilon ^{2}$ in (\ref{h0}). However, the problem under consideration
would remain singular, since the domain $\Omega _{\varepsilon }$ shrinks to
the graph $\Gamma $ as $\varepsilon \rightarrow 0.$ The presence of this
coefficient is convenient, since it makes the spectrum less vulnerable to
changes in $\varepsilon .$ As we shall see, in some important cases the
spectrum of the problem does not depend on $\varepsilon ,$ and the spectrum
will be proportional to $\varepsilon ^{-2}$ if $\varepsilon ^{2}$ in (\ref
{h0}) is omitted. The operator in $L^{2}(\Omega _{\varepsilon })$
corresponding to the problem (\ref{h0}) will be denoted by $H_{\varepsilon
}. $

The goal of this paper is the asymptotic analysis of the spectrum of $%
H_{\varepsilon },$ the resolvent $(H_{\varepsilon }-\lambda )^{-1},$ and
solutions of the corresponding non-stationary problems for the heat and wave
equations as $\varepsilon \rightarrow 0.$ One can expect that $%
H_{\varepsilon }$ is close (in some sense) to a one dimensional operator on
the limiting graph $\Gamma $ with appropriate gluing conditions at the
vertices $v\in V.$ The justification of this fact is not always simple. The
form of the GC in general situation was discovered quite recently in our
previous paper \cite{MV1}.

An important class of domains $\Omega _{\varepsilon }$ are self-similar
domains with only one junction and all the channels being infinite. We shall
call them \textit{spider domains}. Thus, if $\Omega _{\varepsilon }$ is a
spider domain, then there exist a point $\widehat{x}=x(\varepsilon )$ and an
$\varepsilon $-independent domain $\Omega $ such that

\begin{equation}
\Omega _{\varepsilon }=\{(\widehat{x}+\varepsilon x):x\in \Omega \}.
\label{hom}
\end{equation}
Thus, $\Omega _{\varepsilon }$ is the $\varepsilon $-contraction of $\Omega
=\Omega _{1}.$

For the sake of simplicity we shall assume that $\Omega _{\varepsilon }$ is
self-similar in a neighborhood of each junction. Namely, let $%
J_{j(v),\varepsilon }$ be the junction which corresponds to a vertex $v\in V$
of the limiting graph $\Gamma .$ Consider a junction $J_{j(v),\varepsilon }$
and all the channels adjacent to $J_{j(v),\varepsilon }$. If some of these
channels have finite length, we extend them to infinity. We assume that, for
each $v\in V,$ the resulting domain $\Omega _{v,\varepsilon }$ which
consists of the junction $J_{j(v),\varepsilon }$ and the semi-infinite
channels emanating from it is a spider domain. We also assume that all the
channels $C_{j,\varepsilon }$ have the same cross-section $\omega
_{\varepsilon }.\ $This assumption is needed only to make the results more
transparent. From the self-similarity assumption it follows that $\omega
_{\varepsilon }\ $is an $\varepsilon -$homothety of a bounded domain $\omega
\in R^{d-1}$. \

Let $\lambda _{0}<\lambda _{1}\leq \lambda _{2}...$ be eigenvalues of the
negative Laplacian $-\Delta _{d-1}$ in $\omega $ with the BC $B_{0}u=0$ on $%
\partial \omega $ where we put $\varepsilon =1$ in $B_{0},$ and let $%
\{\varphi _{n}(y)\},$ $y\in \omega \in R^{d-1},$\ be the set of
corresponding orthonormal eigenfunctions.\ Then $\lambda _{n}$ are
eigenvalues of $-\varepsilon ^{2}\Delta _{d-1}$ in $\omega _{\varepsilon }$
and $\{\varphi _{n}(y/\varepsilon )\}$ are the corresponding eigenfunctions$%
. $ $\ $In the presence of infinite channels, the spectrum of the operator $%
H_{\varepsilon }$ consists of an absolutely continuous component which
coincides with the semi-bounded interval $[\lambda _{0},\infty )$ and a
discrete set of eigenvalues. The eigenvalues can be located below $\lambda
_{0}$ and can be embedded into the absolutely continuous spectrum$.$ We will
call the point $\lambda =\lambda _{0}$ the threshold since it is the bottom
of the absolutely continuous spectrum or (and) the first point of
accumulation of the eigenvalues as $\varepsilon \rightarrow 0.$ Let us
consider two simplest examples: the Dirichlet problem in a half infinite
cylinder and in a bounded cylinder of the length $l.$ In the first case, the
spectrum of the negative Dirichlet Laplacian in $\Omega _{\varepsilon }$\ is
pure absolutely continuous and has multiplicity $n+1$ on the interval $%
[\lambda _{n},\infty ).$ In the second case the spectrum consists of the set
of eigenvalues $\lambda _{n,m}=\lambda _{n}+\varepsilon ^{2}m^{2}/l^{2},$ $%
n\geq 0,$ $m\geq 1.$

The wave propagation governed by the operator $H_{\varepsilon }$ can be
described in terms of the scattering solutions and scattering matrices
associated to individual junctions of $\Omega _{\varepsilon }.$ The
scattering solutions give information on the absolutely continuous spectrum
and the resolvent for energies in the bulk of the spectrum ($\lambda
>\lambda _{0}$). The spectrum in a small neighbourhood of $\lambda _{0}$ and
below $\lambda _{0}$ is associated with the parabolic equation. However, the
scattering solutions allow us to approximate the operator $H_{\varepsilon },$
$\varepsilon \rightarrow 0,$ by a one dimensional operator on the limiting
graph for all values of $\lambda \geq \lambda _{0}$. In particular, when $%
\lambda \approx \lambda _{0}$ the corresponding GC on the limiting graph are
expressed in terms of the limits of the scattering matrices as $\lambda
\rightarrow \lambda _{0}$.

The plan of the paper is as follows. Next section is devoted to historical
remarks. A more detailed description of the results from our paper \cite{MV1}
on the asymptotic behavior of the scattering solutions ($\lambda >\lambda
_{0}+\delta ,$ $\varepsilon \rightarrow 0$) is given in section 3. In
particular, the GC on the limiting graph are described ($\lambda >\lambda
_{0}+\delta $)$.$ The Green function of the one dimensional problem on the
limiting graph is studied in section 4. The resolvent convergence as $%
\varepsilon \rightarrow 0$ is established in section 5 when $\lambda $ is
near $\lambda _{0}.$ This\ allows us to derive and rigorously justify the GC
for the limiting problem with $\lambda \approx \lambda _{0}$ which where
obtained earlier \cite{MV1} only on a formal level, as the limit as $\lambda
\rightarrow \lambda _{0}$ of the GC with $\lambda >\lambda _{0}.$ A detailed
analysis of these GC is given.

Let $\lambda =\lambda _{0}+O(\varepsilon ^{2}).$ It has been known (see
references in the next section) that for an arbitrary domain $\Omega
_{\varepsilon }$ and the Neumann boundary condition on $\partial \Omega
_{\varepsilon },$ the GC\ on the limiting graph $\Gamma $ is Kirchhoff's
condition. The GC are different for other boundary conditions on $\partial
\Omega _{\varepsilon }.$ It was shown in \cite{MV1} that for generic domains
$\Omega _{\varepsilon }$ and the boundary conditions different from the
Neumann condition$,$ the GC at the vertices of $\Gamma $ are Dirichlet
conditions$.$ It is shown here that, for arbitrary domain $\Omega
_{\varepsilon },$ the GC at each vertex $v$ of the limiting graph has the
following form. For any function $\varsigma $ on $\Gamma ,$ we form a vector
$\varsigma ^{(v)}\ $whose\ components are restrictions of $\varsigma $ to
the edges of $\Gamma $ adjacent to $v.$ The GC at $v,$ with $\lambda $ near $%
\lambda _{0},$ are the Dirichlet condition for some components of the vector
$\widetilde{\varsigma }^{(v)}\ $and the Neumann condition for the remaining
components where $\widetilde{\varsigma }^{(v)}\ $is a rotation of $\varsigma
^{(v)}$.

Note that the resolvent convergence provides the convergence of the discrete
spectrum. In the presence of finite channels $C_{j,\varepsilon },$ the
operator $H_{\varepsilon }$ has a sequence of eigenvalues which converge to $%
\lambda _{0}$ as $\varepsilon \rightarrow 0$ (see the example above). Thus,
these eigenvalues are asymptotically ($\varepsilon \rightarrow 0$) close to
the eigenvalues of the problem where the junctions are replaced by
Dirichlet/Neumann boundary conditions. The final result concerns the inverse
scattering problem. The GC of the limiting problem depend on $\lambda $ if $%
\lambda >\lambda _{0}+\delta .$ A $\lambda $-independent effective potential
is constructed in the last section of the paper which has the same
scattering data as the original problem. This allows one to reduce the
problem in $\Omega _{\varepsilon }$ to a one dimensional problem with $%
\lambda $-independent GC.

\section{Historical remarks.}

Certain problems related to the operator $H_{\varepsilon }$ have been
studied in detail. They concern, directly or indirectly, the spectrum near
the origin for the operator $H_{\varepsilon }$ with the Neumann boundary
condition on $\partial \Omega _{\varepsilon }$, see \cite{EP,ES,FW,
FW1,K,KZ1,KZ2,P1,RS}. The following couple of features distinguish the
Neumann boundary condition. First, only in this case the ground states $%
\varphi _{0}(y/\varepsilon )=1$ on the cross sections of the channels can be
extended smoothly onto the junctions (by 1) to provide the ground state for
the operator $H_{\varepsilon }$ in an arbitrary domain $\Omega _{\varepsilon
}.$ Another important fact, which is valid only in the case of the Neumann
boundary conditions, is that $\lambda _{0}=0.$ Note that an eigenvalue $%
\lambda =\mu $ of the operator $H_{\varepsilon }$ contributes a term of
order $e^{-\frac{\mu t}{\varepsilon ^{2}}}$ to the solutions of the heat
equation in $\Omega _{\varepsilon }.$ The existence of the spectrum in a
small (of order $O(\varepsilon ^{2})$) neighborhood of the origin leads to
the existence of a non-trivial limit, as $\varepsilon \rightarrow 0$, for
the solutions of the heat equation. Solutions of the heat equation with
other boundary conditions are vanishing exponentially as $\varepsilon
\rightarrow 0$.

The GC and the justification of the limiting procedure $\varepsilon
\rightarrow 0$ when $\lambda $ is near $\lambda _{0}=0$ and the Neumann BC
is imposed at the boundary of $\Omega _{\varepsilon }$ can be found in \cite
{FW}, \cite{KZ1}, \cite{KZ2}, \cite{RS}. Typically, the GC at the vertices
of the limiting graph in this case are:$\;$the continuity at each vertex $v$
of both the field and the flow. These GC are called Kirchhoff's GC. The
paper \cite{FW} provides the convergence, as $\varepsilon \rightarrow 0$, of
the Markov process on $\Omega _{\varepsilon }$ to the Markov process on the
limiting graph for more general domains $\Omega _{\varepsilon }$ (the cross
section of a channel can vary). In the case when the shrinkage rate of the
volume of the junctions is lower than the one of the area of the
cross-sections of the guides, more complex, energy dependent or decoupling,
conditions may arise (see \cite{K}, \cite{KZ2}, \cite{EP} for details).

The operator $H_{\varepsilon }$ with the Dirichlet boundary condition on $%
\partial \Omega _{\varepsilon }$ was studied in a recent paper \cite{P}
under conditions that $\lambda $ is near the threshold $\lambda _{0}>0$ and
the junctions are more narrow than the channels. It is assumed there that
the domain $\Omega _{\varepsilon }$ is bounded. Therefore, the spectrum of
the operator (\ref{h0}) is discrete. It is proved that the eigenvalues of
the operator (\ref{h0}) in a small$\ $neighborhood of $\lambda _{0}$ behave
asymptotically, when $\varepsilon \rightarrow 0,$ as eigenvalues of the
problem in the disconnected domain that one gets by omitting the junctions,
separating the channels in $\Omega _{\varepsilon },$ and adding the
Dirichlet conditions on the bottoms of the channels. This result indicates
that the waves do not propagate through the narrow junctions when $\lambda $
is close to the bottom of the absolutely continuous spectrum. A similar
result was obtained in \cite{IT} for the Schr\"{o}dinger operator with a
potential having a deep strict minimum on the graph, when the width of the
walls shrinks to zero. It will be shown in this paper, that the same result
(the GC on the limiting graph is the Dirichlet condition if $H_{\varepsilon
} $ is the operator with the Dirichlet boundary condition on $\partial
\Omega _{\varepsilon }$ and $\lambda \approx \lambda _{0}>0$) is valid for
generic domains $\Omega _{\varepsilon }$ without assumptions on the size of
the junctions.

The asymptotic analysis of the scattering solutions and the resolvent for
operator $H_{\varepsilon }$ with arbitrary boundary conditions on $\partial
\Omega _{\varepsilon }$ and $\lambda $ in the bulk of the absolutely
continuous spectrum ($\lambda >\lambda _{0}+\delta $) was given by us in
\cite{MV}, \cite{MV1}. It was shown there that the GC\ on the limiting graph
can be expressed in terms of the scattering matrices defined by junctions of
$\Omega _{\varepsilon }.$ Formal extension of these conditions to $\lambda
=\lambda _{0}$ leads to the Dirichlet boundary conditions at the vertices of
the limiting graph for generic domains $\Omega _{\varepsilon }.$ Among other
results, we will show here that the asymptotics obtained in \cite{MV}, \cite
{MV1} are valid up to $\lambda =\lambda _{0}.$ $\ $

There is extended literature on the spectrum of the operator $H_{\varepsilon
}$ below the threshold $\lambda _{0}$ (for example, see \cite{DE}-\cite{ExW}
and references therein). We shall not discuss this topic in the present
paper. Important facts on the scattering solutions in networks of thin
fibers can be found in \cite{Pavlov2}, \cite{Pavlov4}.

\section{Scattering solutions.}

We introduce Euclidean coordinates $(t,y)$ in channels $C_{j,\varepsilon }$
chosen in such a way that $t$-axis is parallel to the axis of the channel,
hyperplane $R_{y}^{d-1}$ is orthogonal to the axis, and $C_{j,\varepsilon }$
has the following form in the new coordinates:
\begin{equation*}
C_{j,\varepsilon }=\{(t,\varepsilon y):0<t<l_{j},\text{ }y\in \omega \}.
\end{equation*}

Let us recall the definition of scattering solutions for the problem in $%
\Omega _{\varepsilon }.$ In this paper, we'll need the scattering solutions
only in the case of $\lambda \in (\lambda _{0},\lambda _{1})$.\ Consider the
non-homogeneous problem
\begin{equation}
(-\varepsilon ^{2}\Delta -\lambda )u=f,\text{ \ }x\in \Omega _{\varepsilon };%
\text{ \ \ \ }Bu=0\text{ \ on }\partial \Omega _{\varepsilon }.  \label{a1}
\end{equation}

\begin{definition}
\label{d1}Let $f\in L_{com}^{2}(\Omega _{\varepsilon })$ have a compact
support, and $\lambda _{0}<\lambda <\lambda _{1}$.\ A solution $u$ of (\ref
{a1}) is called outgoing if it has the following asymptotic behavior at
infinity in each infinite channel $C_{j,\varepsilon },$ \ \ $1\leq j\leq m$:
\begin{equation}
u=a_{j}e^{i\frac{\sqrt{\lambda -\lambda _{0}}}{\varepsilon }t}\varphi
_{0}(y/\varepsilon )+O(e^{-\gamma t}),\text{ \ \ }\gamma =\gamma
(\varepsilon ,\lambda )>0.  \label{a2}
\end{equation}
If $\lambda <\lambda _{0},$ a solution $u$ of (\ref{a1}) is called outgoing
if it decays at infinity.

\begin{definition}
\label{d2}Let $\lambda _{0}<\lambda <\lambda _{1}.$ A function $\Psi =\Psi
_{p}^{(\varepsilon )},$ $0\leq p\leq m,$ is called a solution of the
scattering problem in $\Omega _{\varepsilon }$ if
\begin{equation}
(-\Delta -\lambda )\Psi =0,\text{ \ }x\in \Omega _{\varepsilon };\text{ \ \
\ }B\Psi =0\text{ \ on }\partial \Omega _{\varepsilon },  \label{b9}
\end{equation}
and $\Psi $ has the following asymptotic behavior in each infinite channel $%
C_{j,\varepsilon },\ \ 1\leq j\leq m:$%
\begin{equation}
\Psi _{p}^{(\varepsilon )}=\delta _{p,j}e^{-i\frac{\sqrt{\lambda -\lambda
_{0}}}{\varepsilon }t}\varphi _{0}(y/\varepsilon )+t_{p,j}e^{i\frac{\sqrt{%
\lambda -\lambda _{0}}}{\varepsilon }t}\varphi _{0}(y/\varepsilon
)+O(e^{-\gamma t}),  \label{b10}
\end{equation}
where $\gamma =\gamma (\varepsilon ,\lambda )>0,$ and $\delta _{p,j}$ is the
Kronecker symbol, i.e. $\delta _{p,j}=1$ if $p=j,$ $\delta _{p,j}=0$ if $p$ $%
\neq j.$
\end{definition}
\end{definition}

The first term in (\ref{b10}) corresponds to the incident wave (coming
through the channel $C_{p,\varepsilon }$), and all the other terms describe
the transmitted waves. The transmission coefficients $t_{p,j}=t_{p,j}(%
\varepsilon ,\lambda )$ depend on $\varepsilon $ and $\lambda $. The matrix

\begin{equation}
T=[t_{p,j}]  \label{scm}
\end{equation}
is called the scattering matrix.

The outgoing and scattering solutions are defined similarly when $\lambda
\in (\lambda _{n},\lambda _{n+1}).$ In this case, any outgoing solution has $%
n+1$ waves in each channel propagating to infinity with the frequencies $%
\sqrt{\lambda -\lambda _{s}}/\varepsilon ,$ $0\leq s\leq n$. There are $%
m(n+1)$ scattering solutions: the incident wave may come through one of $m$
infinite channels with one of $(n+1)$ possible frequencies. The scattering
matrix has the size $m(n+1)\times m(n+1)$ in this case.

\begin{theorem}
\label{t3}The scattering matrix $T,$ $\lambda >\lambda _{0},$\ $\lambda
\notin \{\lambda _{j}\},$\ is unitary and symmetric ($t_{p,j}=t_{j,p}$).
\end{theorem}

The operator $H_{\varepsilon }$ is non-negative, and therefore the resolvent
\begin{equation}
R_{\lambda }=(H_{\varepsilon }-\lambda )^{-1}:L^{2}(\Omega _{\varepsilon
})\rightarrow L^{2}(\Omega _{\varepsilon })  \label{res}
\end{equation}
is analytic in the complex $\lambda $ plane outside the positive semi-axis $%
\lambda \geq 0.$ Hence, the operator $R_{k^{2}}$ is analytic in $k$ in the
half plane Im$k>0.$ We are going to consider the analytic extension of the
operator $R_{k^{2}}$ to the real axis and the lower half plane. Such an
extension does not exist if $R_{k^{2}}$ is considered as an operator in $%
L^{2}(\Omega _{\varepsilon })$ since $R_{k^{2}}$ is an unbounded operator
when $\lambda =k^{2}$ belongs to the spectrum of the operator $R_{\lambda }.$
However, one can extend $R_{k^{2}}$ analytically if it is considered as an
operator in the following spaces (with a smaller domain and a larger range):
\begin{equation}
R_{k^{2}}:L_{com}^{2}(\Omega _{\varepsilon })\rightarrow L_{loc}^{2}(\Omega
_{\varepsilon }).  \label{b2}
\end{equation}

\begin{theorem}
\label{t1}(1) The spectrum of the operator $H_{\varepsilon }$ consists of
the absolutely continuous component $[\lambda _{0},\infty )$ (if $\Omega
_{\varepsilon }$ has at least one infinite channel) and, possibly, a
discrete set of positive eigenvalues $\lambda =\mu _{j,\varepsilon }$ with
the only possible limiting point at infinity.\ The multiplicity of the a.c.
spectrum changes at points $\lambda =\lambda _{n},$ and is equal to $m(n+1)$
on the interval $(\lambda _{n},\lambda _{n+1})$.

If $\Omega _{\varepsilon }$ is a spider domain, then the eigenvalues $\mu
_{j,\varepsilon }$ $=\mu _{j}$ do not depend on $\varepsilon .$

(2) The operator (\ref{b2}) admits a meromorphic extension from the upper
half plane Im$k>0$ into lower half plane Im$k<0$ with the branch points at $%
k=\pm \sqrt{\lambda _{n}}$ of the second order and the real poles at $k=\pm
\sqrt{\mu _{j,\varepsilon }}$ and, perhaps, at some of the branch points
(see the remark below). The resolvent (\ref{b2}) has a pole at $k=\pm \sqrt{%
\lambda _{n}}$ if and only if the homogeneous problem (\ref{a1}) with $%
\lambda =\lambda $ has a nontrivial solution $u$ such that
\begin{equation}
u=a_{j}\varphi _{n}(y/\varepsilon )+(e^{-\gamma t}),\text{ \ \ \ }x\in
C_{j,\varepsilon },\text{ \ \ }t\rightarrow \infty ,\text{ \ \ }1\leq j\leq
m.  \label{inf}
\end{equation}

(3) If $f\in L_{com}^{2}(\Omega _{\varepsilon }),$ and $k=\sqrt{\lambda }$
is real and is not a pole or a branch point of \ the operator (\ref{b2}),
and $\lambda >\lambda _{0},$ then the problem (\ref{a1}), (\ref{a2}) is
uniquely solvable and the outgoing solution $u$ can be found as the $%
L_{loc}^{2}(\Omega _{\varepsilon })$ limit
\begin{equation}
u=R_{\lambda +i0}f.  \label{b3}
\end{equation}

(4) There exist exactly $m(n+1)\ $ different scattering solutions for the
values of $\lambda \in (\lambda _{n},\lambda _{n+1})$ such that $k=\sqrt{%
\lambda }$ is not a pole of \ the operator (\ref{b2}), and the scattering
solution is defined uniquely after the incident wave is chosen.
\end{theorem}

\textbf{Remark.} The pole of $R_{\lambda }$ at a branch point $\lambda
=\lambda _{n}$ is defined as the pole of this operator function considered
as a function of $z=\sqrt{\lambda -\lambda _{n}}.$

Let us describe the asymptotic behavior of scattering solutions $\Psi =\Psi
_{p}^{(\varepsilon )}$ as $\varepsilon \rightarrow 0,$ $\lambda \in (\lambda
_{0},\lambda _{1}).$ We shall consider here only the first zone of the
absolutely continuous spectrum, but one can find the asymptotics of $\Psi
_{p}^{(\varepsilon )}$ in \cite{MV1} for any $\lambda >\lambda _{0}.$ Note
that an arbitrary solution $u$ of equation (\ref{h0}) in a channel $%
C_{j,\varepsilon }$ can be represented as a series with respect to the
orthogonal basis $\{\varphi _{n}(y/\varepsilon )\}$ of the eigenfunctions of
the Laplacian in the cross-section of $C_{j,\varepsilon }.$ Thus, it can be
represented as a linear combination of the travelling waves
\begin{equation*}
e^{\pm i\frac{\sqrt{\lambda -\lambda _{n}}}{\varepsilon }t}\varphi
_{n}(y/\varepsilon ),\text{ \ \ }\lambda \in (\lambda _{n},\lambda _{n+1}),
\end{equation*}
and terms which grow or decay exponentially along the axis of $%
C_{j,\varepsilon }.$ The main term of small $\varepsilon $ asymptotics of
scattering solutions contains only travelling waves, i.e. functions $\Psi
_{p}^{(\varepsilon )}$ in each channel $C_{j,\varepsilon }$ have the
following form when $\lambda \in (\lambda _{0},\lambda _{1}):$
\begin{equation}
\Psi =\Psi _{p}^{(\varepsilon )}=(\alpha _{p,j}e^{-i\frac{\sqrt{\lambda
-\lambda _{0}}}{\varepsilon }t}+\beta _{p,j}e^{i\frac{\sqrt{\lambda -\lambda
_{0}}}{\varepsilon }t})\varphi _{0}(y/\varepsilon )+r_{p,j}^{\varepsilon },%
\text{ \ \ \ }x\in C_{j,\varepsilon },  \label{psias}
\end{equation}
where
\begin{equation*}
|r_{p,j}^{\varepsilon }|\leq Ce^{-\frac{\gamma d(t)}{\varepsilon }},\text{ \
\ }\gamma >0,\text{ \ \ and \ }d(t)=\min (t,l_{j}-t).
\end{equation*}
The constants $\alpha _{p,j},$ $\beta _{p,j}$ and functions $%
r_{p,j}^{\varepsilon }$ depend on $\lambda $ and $\varepsilon .$ Formula (%
\ref{psias}) can be written as follows
\begin{equation}
\Psi =\Psi _{p}^{(\varepsilon )}=\varsigma \varphi _{0}(y/\varepsilon
)+r_{p}^{\varepsilon },\text{ \ \ \ }|r_{p}^{\varepsilon }|\leq Ce^{-\frac{%
\gamma d(t)}{\varepsilon }},  \label{ps1}
\end{equation}
where the function $\varsigma =\varsigma (t)$ can be considered as a
function on the limiting graph $\Gamma $ which is equal to $\varsigma
_{j}(t),$\ $0<t<l_{j},$ on the edge $\Gamma _{j}$ and satisfies the
following equation:
\begin{equation}
(\varepsilon ^{2}\frac{d^{2}}{dt^{2}}+\lambda -\lambda _{0})\varsigma =0.
\label{greq}
\end{equation}

In order to complete the description of the main term of the asymptotic
expansion (\ref{psias}), we need to provide the choice of constants in the
representation of $\varsigma _{j}$ as a linear combinations of the
exponents. We specify $\varsigma $ by imposing conditions at infinity and
gluing conditions (GC) at each vertex $v$ of the graph $\Gamma .$ Let $%
V=\{v\}$ be the set of vertices $v$ of the limiting graph $\Gamma .$ These
vertices correspond to the free ends of the channels and the junctions in $%
\Omega _{\varepsilon }.$

The conditions at infinity concern only the infinite channels $%
C_{j,\varepsilon },$ $j\leq m.$ They indicate that the incident wave comes
through the channel $C_{p,\varepsilon }$. They have the form:
\begin{equation}
\beta _{p,j}=\delta _{p,j}.  \label{uv}
\end{equation}

The GC at vertices $v$ of the graph $\Gamma $ are universal for all incident
waves and depend on $\lambda $. We split the set $V$ of vertices $v$ of the
graph in two subsets $V=V_{1}\cup V_{2},$ where the vertices from the set $%
V_{1}$ have degree $1$ and correspond to the free ends of the channels, and
the vertices from the set $V_{2}$ have degree at least two and correspond to
the junctions $J_{j,\varepsilon }$. We keep the same BC at $v\in V_{1}$ as
at the free end of the corresponding channel of $\Omega _{\varepsilon }:$%
\begin{equation}
B_{e}\zeta =0\text{ \ \ at }v\in V_{1}.  \label{bce}
\end{equation}
In order to state the GC at a vertex $v\in $ $V_{2}$, we choose the
parametrization on $\Gamma $ in such a way that $t=0$ at $v$ for all edges
adjacent to this particular vertex$.$ The origin ($t=0$) on all the other
edges can be chosen at any of the end points of the edge. Let $d=d(v)\geq 2$%
\ be the order (the number of adjacent edges) of the vertex $v\in V_{2}.$
For any function $\varsigma $ on $\Gamma ,$ we form a vector $\varsigma
^{(v)}=\varsigma ^{(v)}(t)$ with $d(v)$\ components equal to\ the
restrictions of $\varsigma $ on the edges of $\Gamma $ adjacent to $v.$ We
shall need this vector only for small values of $t\geq 0.$ Consider
auxiliary scattering problems for the spider domain $\Omega _{v,\varepsilon
}.$ The domain is formed by the individual junction which corresponds to the
vertex $v,$ and all channels with an end at this junction, where the
channels are extended to infinity if they have a finite length. We enumerate
the channels of $\Omega _{v,\varepsilon }$ according to the order of the
components of the vector $\varsigma ^{(v)}.$ We denote by $\Gamma _{v}$ the
limiting graph defined by $\Omega _{v,\varepsilon }.$ Definitions \ref{d1},
\ref{d2} and Theorem \ref{t1} remain valid for the domain $\Omega
_{v,\varepsilon }.$ In particular, one can define the scattering matrix $%
T=T_{v}(\lambda )$ for the problem (\ref{h0}) in the domain $\Omega
_{v,\varepsilon }.$ Let $I_{v}$ be the unit matrix of the same size as the
size of the matrix $D_{v}(\lambda ).$ The GC at the vertex $v\in V_{2}$ has
the form
\begin{equation}
i\varepsilon \lbrack I_{v}+T_{v}(\lambda )]\frac{d}{dt}\varsigma ^{(v)}(t)-%
\sqrt{\lambda -\lambda _{0}}[I_{v}-T_{v}(\lambda )]\varsigma ^{(v)}(t)=0,%
\text{ \ \ \ }t=0,\text{ \ \ \ }v\in V_{2}.  \label{gc}
\end{equation}
One has to keep in mind that the self-similarity of the spider domain $%
\Omega _{v,\varepsilon }$ implies that $T_{v}=T_{v}(\lambda )$ does not
depend on $\varepsilon .$

\begin{definition}
A family of subsets $l(\varepsilon )$ of a bounded closed interval $l\subset
R^{1}$ will be called thin if, for any $\delta >0,$ there exist\ constants $%
\beta >0$ and $c_{1},$ independent of $\delta $ and $\varepsilon ,$ and $%
c_{2}=$ $c_{2}(\delta ),$ such that $l(\varepsilon )$ can be covered by $%
c_{1}$ intervals of length $\delta $ together with $c_{2}\varepsilon ^{-1}$
intervals of length $c_{2}e^{-\beta /\varepsilon }.$ Note that $%
|l(\varepsilon )|\rightarrow 0$ as $\varepsilon \rightarrow 0.$
\end{definition}

\begin{theorem}
\label{t2}For any bounded closed interval $l\subset (\lambda _{0},\lambda
_{1}),$ there exists $\gamma =\gamma (l)>0$ and a thin family of sets $%
l(\varepsilon )$ such that the asymptotic expansion (\ref{ps1}) holds on all
(finite and infinite) channels $C_{j,\varepsilon }$ uniformly in $\lambda
\in l$ $\backslash $ $l(\varepsilon )$ and $x$ in any bounded region of $%
R^{d}.$ The function $\varsigma $ in (\ref{ps1}) is a vector function on the
limiting graph which satisfies the equation (\ref{greq}), conditions (\ref
{uv}) at infinity, BC (\ref{bce}), and the GC (\ref{gc}).
\end{theorem}

\textbf{Remarks}. 1) For spider domains, the estimate of the remainder is
uniform for all $x\in R^{d}$.

2) The asymptotics stated in Theorem \ref{t2} is valid only outside of a
thin set $l(\varepsilon )$ since the poles of resolvent (\ref{res}) may run
over the interval $(\lambda _{0},\lambda _{1})$ as $\varepsilon \rightarrow
0,$ and the scattering solution may not exist when $\lambda $ is a pole of
the resolvent. These poles do not depend on $\varepsilon $ for spider
domains, and the set $l(\varepsilon )$ is $\varepsilon $-independent in this
case.

Consider a spider domain $\Omega _{v,\varepsilon }$ and scattering solutions
$\Psi =\Psi _{p}^{(\varepsilon )}$ in $\Omega _{v,\varepsilon }$ when $%
\lambda $ belongs to a small neighborhood of $\lambda _{0},$ i.e. $\ $%
\begin{equation}
\Psi =\Psi _{p,v}^{(\varepsilon )}=(\delta _{p,j}e^{-i\frac{\sqrt{\lambda
-\lambda _{0}}}{\varepsilon }t}+t_{p,j}(\lambda )e^{i\frac{\sqrt{\lambda
-\lambda _{0}}}{\varepsilon }t})\varphi _{0}(y/\varepsilon
)+r_{p,j}^{\varepsilon },\text{ \ \ \ }x\in C_{j,\varepsilon },\text{ \ \ }%
|\lambda -\lambda _{0}|\leq \delta ,  \label{lsc1}
\end{equation}
where $r_{p,j}^{\varepsilon }$ decays exponentially as $t\rightarrow \infty $
and $t=t(x)$ is the coordinate of the point $x.$ We define these solutions
for all complex $\lambda $ in the circle $|\lambda -\lambda _{0}|\leq \delta
$ by the asymptotic expansion (\ref{lsc1}) when Im$\lambda \geq 0, \lambda
\neq \lambda _{0},$ and by extending them analytically for other values of $%
\lambda $ in the circle.

\begin{lemma}
\label{l1n}Let $\Omega _{v,\varepsilon }$ be a spider domain. Then there
exist $\delta $ and $\gamma >0$ such that

1) for each $p$ and $|\lambda -\lambda _{0}|\leq \delta $, $\lambda \neq
\lambda _{0}$, the scattering solution exists and is unique,

2) the scattering coefficients $t_{p,j}(\lambda )$ are analytic in $\sqrt{%
\lambda -\lambda _{0}}$ when $|\lambda -\lambda _{0}|\leq \delta ,$

3) the following estimate is valid for the remainder
\begin{equation*}
|r_{p,j}^{\varepsilon }|\leq \frac{C}{|\lambda -\lambda _{0}|}e^{-\frac{%
\gamma t}{\varepsilon }},\text{ \ \ }\gamma >0,\text{ \ \ }\ x\in
C_{j,\varepsilon }.
\end{equation*}
\end{lemma}

This statement can be extracted from the text of our paper \cite{MV1}. Since
it was not stated explicitly, we shall derive it from the theorems above. In
fact, since the spider domain $\Omega _{v,\varepsilon }$ is self-similar, it
is enough to prove this lemma when $\varepsilon =1.$ We omit index $%
\varepsilon $ in $\Omega _{\varepsilon },$ $C_{p,\varepsilon },$ $\Psi
_{p}^{(\varepsilon )}$ when the problem in $\Omega _{\varepsilon }$ is
considered with $\varepsilon =1.$ Let $\alpha _{p}(x)$ be a $C^{\infty }$-
function on $\Omega $ which is equal to zero outside of the channel $C_{p}$
and equal to one on $C_{p}$ when $t>1.$ We look for the solution $\Psi _{p}$
of the scattering problem in the form
\begin{equation*}
\Psi _{p}=\delta _{p,j}e^{-i\sqrt{\lambda -\lambda _{0}}t}\varphi
_{0}(y)\alpha _{p}(x)+u_{p}.\text{ \ \ \ \ \ }\Im\lambda \geq 0,\text{ }%
\lambda \neq \lambda _{0},
\end{equation*}
Then $u_{p}$ is the outgoing solution of the problem
\begin{equation*}
(-\Delta -\lambda )u=f,\text{ \ }x\in \Omega ;\text{ \ \ \ }u=0\text{ \ on }%
\partial \Omega ,
\end{equation*}
where
\begin{equation*}
f=-\delta _{p,j}[2\nabla (e^{i\sqrt{\lambda -\lambda _{0}}t}\varphi
_{0}(y))\nabla \alpha _{p}(x)+e^{i\sqrt{\lambda -\lambda _{0}}t}\varphi
_{0}(y)\Delta \alpha _{p}(x)]\in L_{com}^{2}(\Omega ).
\end{equation*}
From Theorem \ref{t1} it follows that there exists $\delta >0$ such that $%
u_{p}$ exists and is unique when $|\lambda -\lambda _{0}|\leq \delta $, $%
\Im\lambda \geq 0,$ $\lambda \neq \lambda _{0}$, and $u_{p}=R_{\lambda }f$
when $\Im\lambda \geq 0,$ and $u_{p}$ can be extended analytically to the
lower half-plane if $R_{\lambda }$ is understood as in (\ref{b2})$.$ The
function $u_{p}=R_{\lambda }f$ may have a pole at $\lambda =\lambda _{0}$.
In particular, on the cross-sections $t=1$ of the infinite channels $C_{j}, $
the function $u_{p}$ is analytic in $\sqrt{\lambda -\lambda _{0}}$ (when $%
|\lambda -\lambda _{0}|\leq \delta $) with a possible pole at $\lambda
=\lambda _{0}$ . We note that $f=0$ on $C_{j}\cap \{t\geq 1\}.$ We represent
$u_{p}$ there as a series with respect to the basis $\{\varphi _{s}(y)\}.$
This leads to (\ref{lsc1}) and justifies all the statements of the lemma if
we take into account the following two facts: 1) the resolvent (\ref{b2})
can not have a singularity at $\lambda =\lambda _{0}$ of order higher than $%
1/|\lambda -\lambda _{0}|$, since the norm of the resolvent (\ref{res}) at
any point $\lambda $ does not exceed the inverse distance from $\lambda $ to
the spectrum, 2) the scattering coefficients can not have a singularity at $%
\lambda =\lambda _{0}$ due to Theorem \ref{t3}.

The proof of Lemma \ref{l1n} is complete.

\section{Spectrum of the problem on the limiting graph.}

Let us write the inhomogeneous problem on the limiting graph $\Gamma $ which
corresponds to the scattering problem (\ref{greq}), (\ref{uv}), (\ref{bce}),
(\ref{gc}). We shall always assume that the function $f$ in the right-hand
side in the equation below has compact support. Then the corresponding
inhomogeneous problem has the form
\begin{equation}
(\varepsilon ^{2}\frac{d^{2}}{dt^{2}}+\lambda -\lambda _{0})\varsigma =f%
\text{ \ on }\Gamma ,\text{ \ \ \ \ }  \label{n1}
\end{equation}
\begin{equation}
B_{e}\varsigma =0\text{ \ at }v\in V_{1},\text{\ \ \ \ \ \ }i\varepsilon
\lbrack I_{v}+T_{v}(\lambda )]\frac{d}{dt}\varsigma ^{(v)}(0)-\sqrt{\lambda
-\lambda _{0}}[I_{v}-T_{v}(\lambda )]\varsigma ^{(v)}(0)=0\text{ \ \ \ \ at }%
v\in V_{2},  \label{n2}
\end{equation}
\begin{equation}
\varsigma =\beta _{j}e^{i\frac{\sqrt{\lambda -\lambda _{0}}}{\varepsilon }t},%
\text{ \ \ }t>>1,\text{\ \ on the infinite edges }\Gamma _{j_{,}}\text{ \ }%
1\leq j\leq m.  \label{n3}
\end{equation}

This problem is relevant to the original problem in $\Omega _{\varepsilon }$
only while $\lambda <\lambda _{1},$ since more than one mode in each channel
survives as $\varepsilon \rightarrow 0$ when $\lambda >\lambda _{1}.$ The
latter leads to a more complicated problem on the limiting graph (see \cite
{MV1}). We are going to use the problem (\ref{n1})-(\ref{n3}) to study the
spectrum of the operator $H_{\varepsilon }$ when
\begin{equation}
\lambda =\lambda _{0}+\varepsilon ^{2}\mu ,\text{ \ \ }|\mu |<c.  \label{lm1}
\end{equation}
As we shall see later, if $\Omega _{\varepsilon }$ has a channel of finite
length, then the operator $H_{\varepsilon }$ has a sequence of eigenvalues
which are at a distance of order $O(\varepsilon ^{2})$ from the threshold $%
\lambda =\lambda _{0}.$ For example, if $\Omega _{\varepsilon }$ is a finite
cylinder with the Dirichlet boundary condition (see the introduction) these
eigenvalues have the form $\lambda _{0}+\varepsilon ^{2}m^{2}/l^{2},$ $m\geq
1.$ Thus, assumption (\ref{lm1}) allows one to study any finite number of
eigenvalues near $\lambda =\lambda _{0}.$

Let us make a substitution $\lambda =\lambda _{0}+\varepsilon ^{2}\mu $ in (%
\ref{n1})-(\ref{n3}). Condition (\ref{n2}) may degenerate at $\mu =0,$ and
one needs to understand this condition at $\mu =0$ as the limit when $\mu
\rightarrow 0$ after an appropriate normalization which will be discussed
later.

\begin{lemma}
\label{gcl0}. There is an orthogonal projection $P=P(v)$ in $R^{d(v)}$ such
that the problem (\ref{n1})-(\ref{n3}) under condition (\ref{lm1}) can be
written in the form
\begin{equation}
(\frac{d^{2}}{dt^{2}}+\mu )\varsigma =\varepsilon ^{-2}f\text{ \ on }\Gamma ;
\label{n4}
\end{equation}
\begin{equation}
B_{e}\varsigma =0\text{ \ at }v\in V_{1};\text{\ \ \ \ \ \ \ \ }P\varsigma
^{(v)}(0)+O(\varepsilon )\frac{d}{dt}\varsigma ^{(v)}(0)=0,\text{ \ }%
P^{\perp }\frac{d}{dt}\varsigma ^{(v)}(0)+O(\varepsilon )\varsigma
^{(v)}(0)=0\text{ \ \ \ \ at }v\in V_{2},  \label{n5}
\end{equation}
\begin{equation}
\varsigma =\beta _{j}e^{i\sqrt{\mu }t},\text{ \ \ }t>>1,\text{\ \ on the
infinite edges }\Gamma _{j_{,}}\text{ \ }1\leq j\leq m,  \label{n6}
\end{equation}
where $d(v)$ is the order (the number of adjacent edges) of the vertex $v$
and $P^{\perp }$ is the orthogonal complement to $P$.
\end{lemma}

\textbf{Remarks.} 1) The GC\ (\ref{n4}) at $v\in V_{2}$ looks particularly
simple in the eigenbasis of the operators $P,$ $P^{\perp }.$ If $\varepsilon
=0$, then it is the Dirichlet/Neumann GC, i.e after appropriate orthogonal
transformation $\xi ^{v}=C_{v}\tilde{\xi}^{v},$
\begin{equation*}
\tilde{\xi}_{1}^{v}\left( 0\right) =\cdots =\tilde{\xi}_{k}^{v}\left(
0\right) =0,\ \frac{d\tilde{\xi}_{k+1}^{v}}{dt}\left( 0\right) =\cdots =%
\frac{d\tilde{\xi}_{d}^{v}}{dt}\left( 0\right) =0,\text{ \ }k=\text{Rank}P.
\end{equation*}
2) We consider $\mu $ as being a spectral parameter of the problem (\ref{n4}%
)-(\ref{n6}), but one needs to keep in mind that the terms $O(\varepsilon )$
in condition (\ref{n4}) depend on $\mu .$

\textbf{Proof.} Let us recall that the matrix $T_{v}(\lambda )$ is analytic
in $\sqrt{\lambda -\lambda _{0}}$ due to Lemma \ref{l1n}. Theorem \ref{t3}
implies the existence of the orthogonal matrix $C_{v}$ such that $%
D_{v}:=C_{v}^{-1}T(\lambda _{0})C_{v}$ is a diagonal matrix with elements $%
\nu _{s}=\pm 1,$ $1\leq s\leq d(v),$ on the diagonal$.$ In fact, from
Theorem \ref{t3} it follows that, for any $\lambda \in \lbrack \lambda
_{0},\lambda _{1}]$, one can reduce $T(\lambda )$ to a diagonal form with
diagonal elements $\nu _{s}=\nu _{s}(\lambda )$ where $|\nu _{s}|=1$.
Additionally, one can easily show that the matrix $T(\lambda _{0})$ is
real-valued, and therefore, $\nu _{s}=\pm 1$ when $\lambda =\lambda _{0}.$
The statement of the lemma follows immediately from here with $P=\frac{1}{2}%
(I-D_{v})C_{v}^{-1}.$

Consider the Green function $G_{\varepsilon }(\gamma ,\gamma _{0},\mu ),$ $%
\gamma ,\gamma _{0}\in \Gamma ,$ $\gamma _{0}\notin V,$ of the problem (\ref
{n4})-(\ref{n6}) which is the solution of the problem with $\varepsilon
^{-2}f$ replaced by $\delta _{\gamma _{0}}(\gamma ).$\ Here $\delta _{\gamma
_{0}}(\gamma )$ is the delta function on $\Gamma $ supported at the point $%
\gamma _{0}$ which belongs to one of the edges of $\Gamma $ ($\gamma _{0}$
is not a vertex). The Green function $G_{0}(\gamma ,\gamma _{0},\mu )$ is
the solution of (\ref{n4})-(\ref{n6}) with $\varepsilon =0$ in (\ref{n5}).

Let us denote by $P_{0}$ the closure in $L^{2}(\Gamma )$ of the operator $-%
\frac{d^{2}}{dt^{2}}$ defined on smooth functions satisfying (\ref{n5})\
with $\varepsilon =0.$ Note that the conditions (\ref{n5}) with $\varepsilon
=0$ do not depend on $\mu .$ Hence, $P_{0}$ is a self-adjoint operator whose
spectrum consists of an absolutely continuous component $\{\mu \geq 0\} $
(if $\Gamma $ has at least one unbounded edge) and a discrete set $\{\mu
_{j}\}$ of non-negative eigenvalues. Let us denote by\ $B_{c}$\ the disk $%
|\mu |<c$ of the complex $\mu $-plane.

\begin{lemma}
\label{nl2}For any $c>0$ there exist $\varepsilon _{0}>0$ such that

1) the eigenvalues $\{\mu _{s}(\varepsilon )\}$ of the problem (\ref{n4})-(%
\ref{n6}) in the disk $B_{c}$ of the complex $\mu $-plane are located in $%
C_{1}\varepsilon $-neighborhoods of the points $\{\mu _{j}\},$ $%
C_{1}=C_{1}(c),$ and each such neighborhood contains $p$ eigenvalues $\mu
_{s}(\varepsilon )$ with multiplicity taken into account where $p$ is the
multiplicity of the eigenvalue $\mu _{j},$

2) the Green function $G_{\varepsilon }$ exists and is unique when $\mu \in
B_{c}\backslash \{\mu _{s}(\varepsilon )\}$ and has the form
\begin{equation*}
G_{\varepsilon }(\gamma ,\gamma _{0},\mu )=\frac{g(\gamma ,\gamma _{0},\mu
,\varepsilon )}{h(\mu ,\varepsilon )},
\end{equation*}
where $g$ is a continuous function of $\gamma \in \Gamma _{m},$ $\gamma
_{0}\in \Gamma _{n},$ $\mu \in B_{c},$ $\varepsilon \in \lbrack
0,\varepsilon _{0}],$ functions $g$ and $h$ are analytic in $\sqrt{\mu }\ $%
and $\varepsilon ,$ and $d$ has zeros in $B_{c}$ only at points $\mu =\mu
_{s}(\varepsilon )$. Here $\Gamma _{m},$ $\Gamma _{n}\ $are arbitrary edges
of $\Gamma .$
\end{lemma}

\textbf{Proof.} We denote by $t=t(\gamma )$ ($t_{0}=t_{0}(\gamma _{0})$) the
value of the parameter on the edge of $\Gamma $ which corresponds to the
point $\gamma \in \Gamma _{m}$ $(\gamma _{0}\in \Gamma _{n},$ respectively).
We look for the Green function in the form
\begin{equation}
G_{\varepsilon }=\delta _{m,n}\frac{e^{i\sqrt{\mu }|t-t_{0}|}}{2i\sqrt{\mu }}%
+a_{m,n}e^{i\sqrt{\mu }t}+b_{m,n}e^{-i\sqrt{\mu }t},\text{ \ \ \ }\gamma \in
\Gamma _{m},\text{ }\gamma _{0}\in \Gamma _{n},  \label{gep}
\end{equation}
where $\delta _{m,n}$ is the Kronecker symbol, and the functions $a_{m,n},$ $%
b_{m,n}$ depend on $t_{0},\mu ,\varepsilon .$ Obviously, (\ref{n4}) with $%
\varepsilon ^{-2}f$ replaced by $\delta _{\gamma _{0}}(\gamma )$ holds. Let
us fix the edge $\Gamma _{n}$ which contains $\gamma _{0}$. We substitute (%
\ref{gep}) into (\ref{n5}), (\ref{n6}) and get $2N$ equations for $2N$ $\ $%
unknowns $a_{m,n},$ $b_{m,n},$ $1\leq m\leq N,$ $n$ is fixed. The matrix $M$
of this system depends analytically on $\sqrt{\mu },$ $\mu \in B_{c}$ and $%
\varepsilon \in \lbrack 0,\varepsilon _{0}]$. The right-hand side has the
form $c_{1}e^{i\sqrt{\mu }t_{0}}+c_{2}e^{-i\sqrt{\mu }t_{0}},$ where the
vectors $c_{1},$ $c_{2}$ depend analytically on $\sqrt{\mu }$ and $%
\varepsilon .$ This implies all the statements of the lemma if we take into
account that the determinant of $M$ with $\varepsilon =0$ has zeroes at
eigenvalues of the operator $P_{0}.$ The proof is complete.

In order to justify the resolvent convergence of the operator $%
H_{\varepsilon }$ as $\varepsilon \rightarrow 0$ and obtain the asymptotic
behavior of the eigenvalues of the problem ( \ref{h0}) near $\lambda _{0}$
we need to represent the Green function $G_{\varepsilon }$ of the problem on
the graph in a special form. We fix points $\gamma _{i}$ strictly inside of
the edges $\Gamma _{i}.$ These points split $\Gamma $ into graphs $\Gamma
_{v}^{cut}$ which consist of one vertex $v$ and parts of adjacent edges up
to corresponding points $\gamma _{i}.$ If $\Gamma _{v}$ is the limiting
graph which corresponds to the spider domain $\Omega _{v,\varepsilon }$,
then $\Gamma _{v}^{cut}$ is obtained from $\Gamma _{v}$ by cutting its edges
at points $\gamma _{i}.$

When $\varepsilon \geq 0$ is small enough, equation (\ref{n4}) on $\Gamma
_{v}$ has $d(v)$ linearly independent solutions satisfying the condition
from (\ref{n5}) which corresponds to the chosen vertex $v.$ This is obvious
if $\varepsilon =0$ (when the components of the vector $C_{v}^{-1}\varsigma
^{(v)}$ satisfy either the Dirichlet or the Neumann conditions at $v$).
Therefore it is also true for small $\varepsilon \geq 0.$ We denote this
solution space by $S_{v}.$ Let us fix a specific basis $\{\psi _{p,v}(\gamma
,\mu ,\varepsilon )\},$ $1\leq p\leq d(v),$ in $S_{v}$. It is defined as
follows. Let us change the numeration of the edges of $\Gamma $ (if needed)
in such a way that the first $d(v)$ edges are adjacent to $v.$ We also
choose the parametrization on these edges in such a way that $t=0$
corresponds to $v.$ Then
\begin{equation*}
\psi _{p,v}=\delta _{p,j}e^{-i\sqrt{\mu }t}+t_{p,j}(\lambda )e^{i\sqrt{\mu }%
t},\text{ \ \ \ }\gamma \in \Gamma _{j}.
\end{equation*}
Here $t=t(\gamma ),$ $\lambda =\lambda _{0}+\varepsilon ^{2}\mu ,$ $%
t_{p,j}=t_{p,j}^{(v)}$ are the scattering coefficients for the spider domain
$\Omega _{v,\varepsilon }.$ Obviously, $\psi _{p,v}$ satisfies conditions (%
\ref{n5}), and formula (\ref{lsc1}) can be written as
\begin{equation}
\Psi _{p,v}^{(\varepsilon )}=\psi _{p,v}(\gamma ,\mu ,\varepsilon )\varphi
_{0}(y/\varepsilon )+r_{p,j}^{\varepsilon },\text{ \ \ \ }x\in
C_{j,\varepsilon },\text{ \ \ }\gamma =\gamma (x).  \label{ppr}
\end{equation}
where $\gamma (x)\in \Gamma _{j}$ is defined by the cross-section of the
channel $C_{j,\varepsilon }$ through the point $x.$

We shall choose one of the points $\gamma _{j}$ in a special way. Namely, if
$\gamma _{0}\in \Gamma _{n}$ then we chose $\gamma _{n}=\gamma _{0}.$ Then $%
G_{\varepsilon }$ belongs to the solution space $S_{v},$ and from Lemma~\ref
{nl2} we get

\begin{lemma}
The Green function $G_{\varepsilon }$ can be represented on each part $%
\Gamma _{v}^{cut}$ of the graph $\Gamma $\ in the form
\begin{equation}
G_{\varepsilon }(\gamma ,\gamma _{0},\mu )=\frac{1}{h(\mu ,\varepsilon )}%
\sum_{1\leq p\leq d(v)}a_{p,v}\psi _{p,v}(\gamma ,\mu ,\varepsilon ),\text{
\ \ \ \ }\mu \in B_{c},\text{ }\varepsilon \in \lbrack 0,\varepsilon _{0}],
\label{cpn}
\end{equation}
where the function $h$ is defined in Lemma~\ref{nl2} and $%
a_{p,v}=a_{p,v}(\gamma _{0},\mu ,\varepsilon )$ are continuous functions
which are analytic in $\varepsilon $ and $\sqrt{\mu }$
\end{lemma}

\section{Resolvent convergence of the operator $H_{\protect\varepsilon }.$}

We are going to study the asymptotic behavior of the resolvent $R_{\lambda
,\varepsilon }=(H_{\varepsilon }-\lambda )^{-1}$ of the operator $%
H_{\varepsilon }$ when (\ref{lm1}) holds and $\varepsilon \rightarrow 0.$
When $\mu $ is complex, the resolvent $R_{\lambda }$ is understood in the
sense of analytic continuation described in Theorem \ref{t1}. In fact, we
shall study $R_{\lambda }f$ only inside of the channels $C_{j,\varepsilon }$
and under the assumption that the support of $f$ belongs to a bounded region
inside of the channels$.$ We fix finite segments $\Gamma _{j}^{\prime
}\subset \Gamma _{j}$ of the edges of the graph large enough to contain the
points $\gamma _{j}.$ Let $\Gamma ^{\prime }=\cup \Gamma _{j}^{\prime }.$ We
denote by $C_{\varepsilon }^{\prime }=\cup C_{j,\varepsilon }^{\prime }$ the
union of the finite parts $C_{j,\varepsilon }^{\prime }$ of the channels
which shrink to $\Gamma _{j}^{\prime }$ as $\varepsilon \rightarrow 0.$ We
shall identify functions from $L^{2}(C_{\varepsilon }^{\prime })$ \ with
functions from $L^{2}(\Omega _{\varepsilon })$ equal to zero outside $%
C_{\varepsilon }^{\prime }.$ We also omit the restriction operator when
functions on $\Omega _{\varepsilon }$ are considered only on $C_{\varepsilon
}^{\prime }.$

If $f\in L^{2}(C_{\varepsilon }^{\prime }),$ denote
\begin{equation*}
\widehat{f}(\gamma ,\varepsilon )=\frac{<f,\varphi _{0}(y/\varepsilon )>}{%
\left| \left| \varphi _{0}(y/\varepsilon )\right| \right| _{L^{2}}}%
=\int_{\Omega _{\varepsilon }}f\varphi _{0}(y/\varepsilon )dy/\left| \left|
\varphi _{0}(y/\varepsilon )\right| \right| _{L^{2}},\text{ \ \ \ }\gamma
\in \Gamma .
\end{equation*}
We shall use the notation $G_{\varepsilon }$ for the integral operator
\begin{equation*}
G_{\varepsilon }\zeta (\gamma )=\int_{\Gamma }G_{\varepsilon }(\gamma
,\gamma _{0},\mu )\zeta (\gamma _{0})d\gamma _{0},\text{ \ \ \ }\gamma \in
\Gamma .
\end{equation*}

\begin{theorem}
\label{tr}Let (\ref{lm1}) hold. Then for any disk $B_{c},$ there exist $%
\varepsilon _{0}=\varepsilon _{0}(c)$\ and a constant $C<\infty $ such that
the function
\begin{equation*}
R_{\lambda ,,\varepsilon }f=(H_{\varepsilon }-\lambda )^{-1}f,\text{ \ \ \ }%
f\in L^{2}(C_{\varepsilon }^{\prime }),
\end{equation*}
is analytic in $\sqrt{\mu }$ when $\mu \in B_{c}\backslash O(\varepsilon )$,
where $O(\varepsilon )$ is $C\varepsilon $-neighborhood of the set $\{\mu
_{j}\},$ and has the form
\begin{equation*}
||R_{\lambda ,\varepsilon }f-\varphi _{0}(y/\varepsilon )G_{\varepsilon }%
\widehat{f}(\gamma ,\varepsilon )||_{L^{2}(C_{\varepsilon }^{\prime })}\leq
C\varepsilon ||f||_{L^{2}(C_{\varepsilon }^{\prime })}.
\end{equation*}
\end{theorem}

\textbf{Remarks.} 1) The points $\mu _{j}$ were introduced above as
eigenvalues of the problem (\ref{n4})-(\ref{n6}) on the graph with $%
\varepsilon =0.$ The GC in this case are the Dirichlet and Neumann
conditions for the components of the vector $C_{v}^{-1}\varsigma ^{(v)}.$
Obviously, these points are also eigenvalues of the operator $H_{\varepsilon
}$ with the junctions of $\Omega _{\varepsilon }$ replaced by the same
Dirichlet/Neumann conditions on the edges of the channels adjacent to the
junctions.

2) The resolvent convergence stated in the theorem implies the convergence,
as $\varepsilon \rightarrow 0,$ of eigenvalues of operator $H_{\varepsilon }$
to $\{\mu _{j}\}.$ We could not guarantee the fact that the eigenvalues of
the problem on the graph are real (see Lemma \ref{nl2}). Of course, they are
real for operator $H_{\varepsilon }.$

\textbf{Proof.} We construct an approximation $K_{\lambda ,\varepsilon }$ to
the resolvent $R_{\lambda ,\varepsilon }=(H_{\varepsilon }-\lambda )^{-1}$
for $f\in L^{2}(C_{\varepsilon }^{\prime }).$ We represent $%
L^{2}(C_{\varepsilon }^{\prime })$ as the orthogonal sum
\begin{equation*}
L^{2}(C_{\varepsilon }^{\prime })=L_{0}^{2}(C_{\varepsilon }^{\prime
})+L_{1}^{2}(C_{\varepsilon }^{\prime }),
\end{equation*}
where functions from $L_{0}^{2}(C_{\varepsilon }^{\prime })$ have the form $%
h(\gamma )\varphi _{0}(y/\varepsilon ),$ $\gamma \in \Gamma ^{\prime },$ and
functions from $L_{1}^{2}(C_{\varepsilon }^{\prime })$ on each cross-section
of the channels are orthogonal to $\varphi _{0}(y/\varepsilon ).$ Here and
below the point $\gamma =\gamma (x)\in \Gamma $ is defined by the
cross-section of the channel through $x.$ We put $\gamma _{0}=\gamma (x_{0}),
$ i.e. $\gamma _{0}$ is the point on the graph defined by the cross-section
of the channel through $x_{0}.$

Consider the operator
\begin{equation*}
K_{\lambda ,\varepsilon }:L^{2}(C_{\varepsilon }^{\prime })\rightarrow
L^{2}(\Omega _{\varepsilon })
\end{equation*}
with kernel $K_{\lambda ,\varepsilon }(x,x_{0})$ defined as follows:
\begin{equation*}
K_{\lambda ,\varepsilon }(x,x_{0})=\sum_{v\in V}\frac{1}{h(\varepsilon ,\mu )%
}\sum_{1\leq p\leq d(v)}a_{p,v}\widehat{\Psi }_{p,v}^{(\varepsilon
)}(x,x_{0}),\text{ \ \ \ \ \ }x_{0}\in C_{\varepsilon }^{\prime },\text{ }%
x\in \Omega _{\varepsilon }.
\end{equation*}
Here \ $h(\mu ,\varepsilon )$ and $a_{p,v}=a_{p,v}(\gamma _{0},\mu
,\varepsilon )$ are functions defined in (\ref{cpn}), and $\widehat{\Psi }%
_{p,v}^{(\varepsilon )}$ are defined by the scattering solutions $\Psi
_{p,v}^{(\varepsilon )}$ of the problem in the spider domain $\Omega
_{v,\varepsilon }$ in the following way. Let $\Omega _{v,\varepsilon }^{0}$
be the part of the spider domain $\Omega _{v,\varepsilon }$ which consists
of the junction and parts of the adjacent channels up to the cylinders $%
C_{j,\varepsilon }^{\prime }.$ Let $\Omega _{v,\varepsilon }^{\prime }$ ($%
\Omega _{v,\varepsilon }^{1}$) be a bigger domain which contains
additionally the parts of the cylinders $C_{j,\varepsilon }^{\prime }$ up to
the cross-sections which correspond to points $\gamma _{j}$ (the whole
cylinders $C_{j,\varepsilon }^{\prime }$, respectively)$.$ We put $\widehat{%
\Psi }_{p,v}^{(\varepsilon )}=\Psi _{p,v}^{(\varepsilon )}$ in $\Omega
_{v,\varepsilon }^{0}.$ We split the scattering solutions $\Psi
_{p,v}^{(\varepsilon )}$ in the cylinders $C_{j,\varepsilon }^{\prime }$
into the sum of two terms. The first term contains the main modes $\varphi
_{0}(y/\varepsilon )e^{\pm i\sqrt{\mu }t},$ and the second one is orthogonal
to $\varphi_{0}(y/\varepsilon )$ in each cross-section. We multiply the
first term by the function $\theta _{v}(x,x_{0})$ equal to one on $\Omega
_{v,\varepsilon }^{\prime }$ and equal to zero everywhere else on $\Omega
_{\varepsilon }.$ We multiply the second term by an infinitely smooth
cut-off function $\eta _{v}(x)$ equal to one on $\Omega _{v,\varepsilon
}^{0} $ and equal to zero on $\Omega_{\varepsilon }$ outside $\Omega
_{v,\varepsilon }^{1}.$ In other terms,
\begin{equation}
\widehat{\Psi }_{p,v}^{(\varepsilon )}(x,x_{0})=\theta _{v}(x,x_{0})\Psi
_{p,v}^{(\varepsilon )}+(\eta _{v}(x)-\theta
_{v}(x,x_{0}))r_{p,j}^{\varepsilon },  \label{ph}
\end{equation}
where $r_{p,j}^{\varepsilon }$ is defined in (\ref{ppr}).

Recall that the representation $\Gamma =\cup \Gamma _{v}^{cut}$ depends on
the choice of points $\gamma _{s}\in \Gamma _{s}^{\prime }\subset \Gamma
_{s}.$ All these points are fixed arbitrarily except one: if $\gamma _{0}\in
\Gamma _{j}$ then $\gamma _{j}$ is chosen to be equal to $\gamma _{0}.$ This
is the reason why $\theta _{v}$ depends on $x_{0}$ and $\eta _{v}$ is $x_{0}$%
-independent.

We look for the parametrix (almost resolvent) of the operator $%
H_{\varepsilon }$, when (\ref{lm1}) holds and $f\in L^{2}(C_{\varepsilon
}^{\prime })$, in the form
\begin{equation*}
F_{\lambda ,\varepsilon }=K_{\lambda ,\varepsilon }P_{0}+R_{\lambda
,\varepsilon }P_{1},
\end{equation*}
where $P_{0},$ $P_{1}$\ are projections on $L_{0}^{2}(C_{\varepsilon
}^{\prime })$ and $L_{1}^{2}(C_{\varepsilon }^{\prime }),$\ respectively. It
is not difficult to show that $||R_{\lambda ,\varepsilon
}P_{1}||=O(\varepsilon ^{2})$ and $H_{\varepsilon }F_{\lambda ,\varepsilon
}=I+F_{\lambda ,\varepsilon },$ where $||F_{\lambda ,\varepsilon
}||=O(\varepsilon ).$ This implies that $R_{\lambda ,\varepsilon
}=K_{\lambda ,\varepsilon }P_{0}+O(\varepsilon ).$ The latter, together with
Lemma \ref{l1n}, justifies Theorem \ref{tr}.

\section{The GC\ at $\protect\lambda $ near the threshold $\protect\lambda %
_{0}$.}

Theorem \ref{tr} and the remarks following the theorem indicate that the GC
at each vertex when $\lambda -\lambda _{0}=O(\varepsilon ^{2})$ is the
Dirichlet/Neumann condition, i.e. the junctions of $\Omega _{\varepsilon }$
can be replaced by $k(v)$ Dirichlet and $d(v)-k(v)$ Neumann conditions at
the edges of the channels adjacent to the junctions \ (after an appropriate
orthogonal transformation). We are going to specify the choice between the
Dirichlet and Neumann conditions. First, we would like to make four important

\textbf{Remarks. }1) Classical Kirchhoff's GC corresponds to $k=d-1.$

2) For any domain $\Omega _{\varepsilon }$ under consideration, if $\lambda
-\lambda _{0}=O(\varepsilon ^{2})$ and the Neumann boundary condition is
imposed on $\partial \Omega _{\varepsilon }$ ($\lambda _{0}=0$ in this case)
then the GC on the limiting graph is Kirchhoff's condition (see section 2).

3) It was proven in \cite{MV1} that if $\lambda -\lambda _{0}=O(\varepsilon
^{2})$ and the boundary condition on $\partial \Omega _{\varepsilon }$ is
different from the Neumann condition, then the GC on the limiting graph is
the Dirichlet condition ($k=d$) for generic domains $\Omega _{\varepsilon }$%
. An example at the end of the next section illustrates this fact.

4) The theorem below states that Kirchhoff's GC\ condition on the limiting
graph appears in the case of arbitrary boundary conditions on $\partial
\Omega _{\varepsilon }$, if the operator $H_{\varepsilon }$ has a ground
state at $\lambda =\lambda _{0}.$ The ground state at $\lambda =\lambda
_{0}=0$ exists for an arbitrary domain $\Omega _{\varepsilon }$, if the
Neumann boundary condition is imposed on $\partial \Omega _{\varepsilon }.$
The ground state at $\lambda =\lambda _{0}$ does not exist for generic
domains $\Omega _{\varepsilon }$ in the case of other boundary conditions
(see \cite{MV1}).

Note that the GC is determined by the scattering matrix in the spider domain
$\Omega _{v,\varepsilon },$ and this matrix does not depend on $\varepsilon
. $ Thus, when the GC is studied, it is enough to consider a spider $%
\varepsilon $-independent domain $\Omega =\Omega _{v,1}.$ We shall omit the
indices $v$ and $\varepsilon $ in $H_{\varepsilon }$, $C_{j,\varepsilon }$
when $\varepsilon =1.$\

\begin{definition}
A ground state of the operator $H$ in a spider domain $\Omega $ at $\lambda
=\lambda _{0}$ is the function $\psi _{0}=\psi _{0}\left( x\right) $, which
is bounded, strictly positive inside $\Omega $, satisfies the equation $%
\left( -\Delta -\lambda _{0}\right) \psi _{0}=0$ in $\Omega$, and the
boundary condition on $\partial \Omega ,$ and has the following asymptotic
behavior at infinity
\begin{equation}
\psi _{0}\left( x\right) =\varphi _{0}\left( y\right) [\rho _{j}+o\left(
1\right) ],\text{ \ }x\in C_{j},\text{ \ }|x|\rightarrow +\infty ,
\label{1b1}
\end{equation}
where $\rho _{j}>0$ and $\varphi _{0}$ is the ground state of the operator
in the cross-sections of the channels$.$
\end{definition}

Let us stress that we assume the strict positivity of $\rho _{j}.$

Let's consider the parabolic problem in a spider domain $\Omega
_{\varepsilon },$
\begin{equation}
\frac{\partial u_{\varepsilon }}{\partial \tau }=\Delta u_{\varepsilon },%
\text{ \ }u_{\varepsilon }\left( 0,x\right) =\varphi \left( \gamma \right)
\varphi _{0}(\dfrac{y}{\varepsilon }),\text{ \ }u_{\varepsilon }\left( \tau
,x\right) |_{\partial \Omega _{\varepsilon }}=0,  \label{a1a}
\end{equation}
where $\gamma =\gamma (x)\in \Gamma $ is defined by the cross-section of the
channel through the point $x$, function $\varphi $ is continuous, compactly
supported with a support outside of the junctions, and depends only on the
longitudinal (''slow'') variable $t\ $on each edge $\Gamma _{j}\subset
\Gamma .$ We shall denote the coordinate $t\ $on $C_{j}$ and $\Gamma _{j}$
by $t_{j}.$ Let $\omega ^{\prime }$ be a compact in the cross-section $%
\omega $ of the channels $C_{j}.$

\begin{theorem}
. \label{1B'} Let $\Omega $ be a spider domain, the Dirichlet or Robin
boundary condition be imposed at $\partial \Omega ,$ and let the operator $H$
have a ground state at $\lambda =\lambda _{0}.$ Then asymptotically, as $%
\varepsilon \rightarrow 0$, the solution of the parabolic problem (\ref{a1a}%
) in $\Omega _{\varepsilon }$ has the following form
\begin{equation*}
u_{\varepsilon }(\tau ,x)=e^{-\frac{\lambda _{0}\tau }{\varepsilon ^{2}}%
}w_{\varepsilon }(\tau ,x)\varphi _{0}(\dfrac{y}{\varepsilon }),
\end{equation*}
where the function $w_{\varepsilon }$ converges uniformly in any region of
the form $0<c^{-1}<\tau <c,$ $t_{j}(x)>\delta >0,$ $y\in \omega ^{\prime }$
to a function $w(\tau ,\gamma )$ on the limiting graph $\Gamma $ which
satisfies the relations
\begin{equation}
\frac{\partial w}{\partial \tau }=\frac{\partial ^{2}w}{\partial t^{2}}\text{
\ \ on }\Gamma ;\text{ \ }w\text{ is\ continuous at the vertex, }%
\sum\limits_{j=1}^{d}\rho _{j}\dfrac{\partial w}{\partial t_{j}}\left(
0\right) =0.  \label{www}
\end{equation}
\end{theorem}

\textbf{Remarks}. 1) Let's note that under the ground state condition,
operator $H_{\varepsilon }$ has no eigenvalues below $\lambda _{0}.$
Otherwise, the eigenfunction with the eigenvalue $\lambda _{\min }<\lambda
_{0}$ must be orthogonal to the ground state $\psi _{0}\left( x\right),$ and
this contradicts the positivity of both functions.

2) The eigenvalues below $\lambda _{0}$ can exist if $H$ does not have the
ground state at $\lambda _{0}.$ For instance, they definitely exist if one
of the junctions is ''wide enough'' (in contrast to the O. Post condition
\cite{P}). The solution $u_{\varepsilon }\left( \tau ,x\right) $ in this
case has asymptotics different from the one stated in Theorem \ref{1B'}. In
particular, if the function $\varphi $ (see (\ref{a1a})) is positive, then
\begin{equation*}
\varepsilon ^{2}\ln u_{\varepsilon }\left( \tau ,x\right) \rightarrow
\lambda _{\min },\text{ \ \ }\tau \rightarrow \infty .
\end{equation*}
What is more important, the total mass of the heat energy in this case is
concentrated in an arbitrarily small, as $\varepsilon \rightarrow 0,$
neighborhood of the junctions. The limiting diffusion process on $\Gamma $
degenerates.

\textbf{Proof. }For simplicity,\textbf{\ }we shall assume that the Dirichlet
boundary condition is imposed on $\partial \Omega _{\varepsilon }.$
Obviously, the function $\psi _{0}\left( \frac{x}{\varepsilon }\right) $ is
the ground state in the spider domain $\Omega _{\varepsilon }$. In
particular,
\begin{equation*}
\varepsilon ^{2}\Delta \psi _{0}+\lambda _{0}\psi _{0}=0;\text{ \ }\psi
_{0}\left( \frac{x}{\varepsilon }\right) =\varphi _{0}(\frac{y}{\varepsilon }%
)[\rho _{j}+o\left( 1\right) ],\text{ \ }x\in C_{j,\varepsilon },\text{ \ }%
|x|\rightarrow +\infty ;\text{ \ \ \ }\psi _{0}|_{\partial \Omega
_{\varepsilon }}=0\text{\ .}
\end{equation*}

Put $u_{\varepsilon }\left( \tau ,x\right) =\psi _{0}\left( \dfrac{x}{%
\varepsilon }\right) e^{-\frac{\lambda _{0}\tau }{\varepsilon ^{2}}%
}w_{\varepsilon }\left( \tau ,x\right) $. Then
\begin{eqnarray}
\frac{\partial w_{\varepsilon }}{\partial \tau } &=&\Delta w_{\varepsilon }+%
\frac{2}{\varepsilon }\nabla \left( \ln \psi _{0}\left( \frac{x}{\varepsilon
}\right) \right) \cdot \nabla w_{\varepsilon },\text{ \ \ }  \notag \\
w_{\varepsilon }\left( 0,x\right) &=&\varphi \left( \gamma \right) \theta
\left( \frac{x}{\varepsilon }\right) ,\text{ \ }\theta \left( \frac{x}{%
\varepsilon }\right) =\frac{1}{\rho _{j}}+o(1),\text{ \ }x\in C_{j},\text{\
\ \ }|x|\rightarrow +\infty .  \label{a2a}
\end{eqnarray}

We look for bounded solutions $u_{\varepsilon },$ $w_{\varepsilon }$\ of the
parabolic problems. We do not need to impose boundary conditions on $%
\partial \Omega _{\varepsilon \text{ }}$ on the function $w_{\varepsilon }$
since the boundedness of $w_{\varepsilon }$ implies that $u_{\varepsilon }=0$
on $\partial \Omega _{\varepsilon }.$ The parabolic problem (\ref{a2a}) has
a unique bounded solution (without boundary conditions on $\partial \Omega
_{\varepsilon }$) since $\nabla _{z}\ln \psi _{0}\left( \cdot \right) $ is
growing near $\partial \Omega _{\varepsilon \text{ }}.$ This growth of the coefficient in (\ref{a2a}) does
not allow the heat energy (or diffusion) to reach $\partial \Omega
_{\varepsilon }$. The fundamental solution $q_{\varepsilon }=q_{\varepsilon
}\left( \tau ,x_{0},x\right) $ of the problem (\ref{a2a}) exists, is unique,
and $\int_{\Omega _{\varepsilon \text{ }}}q_{\varepsilon }\left( \tau
,x_{0},x\right) dx=1.$ This fundamental solution is the transition density
of the Markov diffusion process $X_{\tau }^{(\varepsilon )}=(T_{\tau
}^{(\varepsilon )},Y_{\tau }^{(\varepsilon )})$ in $\Omega _{\varepsilon }$
with the generator $\mathcal{\tilde{H}}_{\varepsilon }=\Delta +\dfrac{2}{%
\varepsilon }\left( \nabla \ln \psi _{0}\left( \dfrac{x}{\varepsilon }%
\right) ,\nabla \right) $ .

Let $\mathcal{\tilde{H}=\tilde{H}}_{\varepsilon }$, and $q=q_{\varepsilon }$
when $\varepsilon =1.$ The coefficients of the operator $\mathcal{\tilde{H}}$
are singular at the boundary of the domain. However, the transition density $%
q\left( \tau ,x_{0},x\right) $ is not vanishing inside $\Omega .$ To be more
exact, the D\"{o}blin condition holds, i.e. for any compact $\omega ^{\prime
}\subset \omega ,$ there exist $\delta >0$ such that for any channel $C_{j}$
the following estimate holds
\begin{equation*}
q(\tau ,x_{0},x)>\delta \text{ \ when \ }T\geq \tau \geq 1,\text{ }x=(t,y),%
\text{ }x_{0}=(t,y_{0}),\text{ }y,\text{ }y_{0}\in \omega ^{\prime }.
\end{equation*}

The operator $\mathcal{\tilde{H}}=\Delta +2\left( \nabla \ln \psi
_{0}(x),\nabla \right) $ has a unique (up to normalization) invariant
measure. This measure has the density $\pi \left( x\right) =\psi
_{0}^{2}\left( x\right) .$ In fact, if $\mathcal{\tilde{H}=}\Delta +\left(
\nabla A,\nabla \right) $ then $\mathcal{\tilde{H}}^{\ast }=\nabla -\left(
\nabla A,\nabla \right) -\left( \Delta A\right) $, and one can easily check
that $\mathcal{\tilde{H}}^{\ast }e^{A\left( x\right) }=0.$ If we put now $%
A\left( x\right) =2\ln \psi _{0}\left( x\right) $, we get $\mathcal{\tilde{H}%
}^{\ast }\pi \left( x\right) =0.$

When $t>\delta _{0}>0$ and $\varepsilon \rightarrow 0$ the transversal
component $Y_{\tau }^{(\varepsilon )}$ and the longitudinal component $%
T_{\tau }^{(\varepsilon )}$ of the diffusion process in $\Omega
_{\varepsilon }$ are asymptotically independent. The transversal component $%
Y_{\tau }^{(\varepsilon )}$ oscillates very fast and has asymptotically ($%
\varepsilon \rightarrow 0$) invariant measure $\dfrac{1}{\varepsilon }%
\varphi _{0}^{2}\left( \dfrac{x}{\varepsilon }\right) $. The latter follows
from the D\"{o}blin condition. The longitudinal component has a constant
diffusion with the drift which is exponentially small (of order $O(e^{-\frac{%
\gamma }{\varepsilon ^{2}}}),$ $\gamma >0$) outside any neighborhood of the
junction.

Under conditions above, one can apply (with minimal modifications) the
fundamental averaging procedure by Freidlin-Wentzel (see \cite{FW}) which
leads to the convergence (in law on each compact interval in $\tau $) of the
distribution of the process $X_{\tau }^{(\varepsilon )}$ to the distribution
of the process on $\Gamma $ with the generator $\dfrac{d^{2}}{dt^{2}\text{ }}
$ on the space of functions on $\Gamma $ smooth outside of the vertex and
satisfying the appropriate GC. The GC\ are defined by the limiting invariant
measure. This limiting measure on $\Gamma $ is equal (up to a normalization)
to $\rho _{j}$ on edges $\Gamma _{j}.$ This leads to the GC\ (\ref{www}) of
the generalized Kirchhoff form. The proof is complete.

\begin{theorem}
. \label{1B} Let operator $H$ in a spider domain $\Omega $ with the
Dirichlet or Robin condition at $\partial \Omega $ have a ground state at $%
\lambda =\lambda _{0},$ and let $\lambda =\lambda _{0}+O(\varepsilon ^{2}).$
Then the GC (\ref{gc}) has the generalized Kirchhoff form: $\zeta $ is
continuous at the vertex $v$ and $\sum_{j=1}^{d}\rho _{j}\frac{d\zeta }{%
dt_{j}}\left( 0\right) =0.$
\end{theorem}

This statement follows immediately from Theorem \ref{1B'} since it is
already established that the GC has the Dirichlet/Neumann form.

\section{ Effective potential.}

As it was already mentioned earlier, the GC (\ref{gc}) is $\lambda $%
-dependent. The following result allows one to reduce the original problem
in $\Omega _{\varepsilon }$ to a Schr\"{o}dinger equation on the limiting
graph with arbitrary $\lambda $-independent GC\ and a $\lambda $-independent
matrix potential. The potential depends on the choice of the GC. Only the
lower part of the a.c. spectrum $\lambda _{0}\leq \lambda \leq \lambda _{1}$
will be considered. It is assumed below that $\varepsilon =1,$ and the index
$\varepsilon $ is omitted everywhere.

Let $T\left( \lambda \right) ,\lambda \in \lbrack \lambda _{0},\lambda
_{1}], $ be the scattering matrix for a spider domain $\Omega $, let $%
\lambda _{-N}\leq \lambda _{-N+1}\leq \cdots \leq \lambda _{-1}<\lambda _{0}$
be the eigenvalues of the discrete spectrum of $H$ below the threshold $%
\lambda _{0} $. The function $T\left( \lambda \right) $ has analytic
extension into the complex plane with the cut along $[0,\infty ).$ It has
poles at $\lambda =\lambda _{-N},...,$ $\lambda _{-1}$. Let $m_{-N},\cdots
,m_{-1}$ be the corresponding residues (Hermitian $d\times d$ matrices).
These residues contain complete information on the multiplicity of $\lambda
_{j},$ $j=-N,\cdots -1$ and on the exponential asymptotics of the
eigenfunctions $\psi _{j}\left( x\right) ,$ $|x|\rightarrow +\infty $.

Theorem \ref{t2} allows one to reduce the problem in $\Omega _{\varepsilon }$
to an equation for a function $\zeta $ on the limiting graph $\Gamma $ with
appropriate GC at the vertex. Consider the vector $\psi :=\zeta ^{(v)}=\zeta
^{(v)}(t),$ $t\geq 0$, whose components are the restrictions of $\zeta $ to
the edges of $\Gamma .$ Note that the GC were formulated through the vector $%
\zeta ^{(v)}.$ Now we would also like to treat the equation for the function
$\zeta $ on $\Gamma $ as a linear system for the vector $\zeta ^{(v)}$ on
the half axis $t\geq 0.$

\begin{theorem}
\label{1C}. There exists an effective fast decreasing matrix $d\times d$
potential $V\left( t\right) $ such that $V\left( t\right) =V^{\ast }\left(
t\right) ,$ and the problem
\begin{equation}
-\psi ^{\prime \prime }+[V\left( t\right) -\lambda _{0}I]\psi =\lambda \psi ,%
\text{ \ }t\geq 0,\text{ \ }\psi \left( 0\right) =0  \label{lll}
\end{equation}
has the same spectral data on the interval $(-\infty ,\lambda _{1})$ as the
original problem in $\Omega .$\ The latter means that the scattering matrix $%
S(\lambda )$ of the problem (\ref{lll}) coincides with $T\left( \lambda
\right) \ \ $on the interval $[\lambda _{0},\lambda _{1}]$, and the poles
and residues of $S(\lambda )$ and $T\left( \lambda \right) \ $are equal.
\end{theorem}

\textbf{Remarks.} 1) The potential is defined not uniquely.

2) The Dirichlet condition $\psi \left( 0\right) =0$ can be replaced by any
fixed GC, say the Kirchhoff one (of course, with the different effective
potential).

3) Different effective potentials appeared when explicitly solvable models
were studied in our paper \cite{MV}.

\textbf{Proof.} This statement is a simple corollary of the inverse spectral
theory by Agranovich and Marchenko for 1-D matrix Schr\"{o}dinger operators
\cite{AM}. One needs only to show that $T\left( \lambda \right) $\ can be
extended to the semiaxis $\lambda >\lambda _{1}$ in such a way that the
extension will satisfy all the conditions required by the
Agranovich-Marchenko theory.

\textbf{Example to the statements of Lemma \ref{gcl0} and Theorem \ref{1B}.}
Consider the Schr\"{o}dinger operator $H=-\frac{d^{2}}{dt^{2}}+v\left(
t\right) $ on the whole axis with a potential $v\left( t\right)$ compactly
supported on $\left[ -1,1\right]$. This operator may serve as a simplified
version of the operator (\ref{lll}). The simplest explicitly solvable model
from \cite{MV} also leads to the operator $H$. The GC at $t=0$\ for this
explicitly solvable model are determined by the limit, as $\varepsilon
\rightarrow 0$, of the solution of the equation $H_{\varepsilon }\psi
_{\varepsilon }=f$, where $H_{\varepsilon }=-\frac{d^{2}}{dt^{2}}%
+\varepsilon ^{-2}v\left( \frac{t}{\varepsilon }\right) $ and $f$ is
compactly supported and vanishing in a neighbourhood of $t=0.$ The solution $%
\psi _{\varepsilon }$ is understood as $L_{loc}^{2}$ limit of $%
(H_{\varepsilon }+i\mu )^{-1}f\in L^{2}$ as $\mu \rightarrow +0.$

Of course, $\lambda =0$ is the bottom of the a.c. spectrum for $H$. If
operator $H$ does not have negative eigenvalues, then the equation $H\psi =0$
has \ a unique (up to a constant factor) positive solution $\psi _{0}\left(
t\right) $, which is not necessarily bounded. If this solution is linear
outside $\left[ -1,1\right]$, then the limiting GC are the Dirichlet ones.
This case is generic. If this solution is constant on one of the semiaxis,
then the GC are the Dirichlet/Neumann conditions. Finally, if $\psi
_{0}\left( t\right) =\rho _{\pm }$ for $\pm t\geq 1$, then we have the
situation of Theorem \ref{1B}: the ground state and the generalized
Kirchhoff's GC.

One can get a nontrivial Kirchhoff's condition even in the case when $H$ has
a negative spectrum. It is sufficient to assume that $\lambda =0$ is the
eigenvalue (but not the minimal one) of the Neumann spectral problem for $H$
on $\left[ -1,1\right] .$


\bigskip

\bigskip

\begin{thebibliography}{99}
\bibitem{AM}  Z. S. Agranovich, V. A. Marchenko, The inverse problem of
scattering theory, 1963, Gordon and Breach Publishers, New-York.

\bibitem{IT}  G. Dell'Antonio, L. Tenuta, Quantum graphs as holonomic
constraints, J. Math. Phys., 47 (2006), pp 072102:1-21.

\bibitem{DE}  P. Duclos, P. Exner, Curvature-induced bound states in quantum
waveguides in two and three dimensions, Rev. Math. Phys., 7 (1995), pp
73-102.

\bibitem{DE3}  P. Duclos, P. Exner, P. Stovicek, Curvature-induced
resonances in a two-dimensional Dirichlet tube, Ann. Inst. H. Poincare 62
(1995), 81-101

\bibitem{EP}  P. Exner, O. Post, Convergence of spectra of graph-like thin
manifolds,, J. Geom. Phys., 54 (2005), 77-115.

\bibitem{ES}  P. Exner , P. \v{S}eba, Electrons in semiconductor
microstructures: a challenge to operator theorists, in Schr\"{o}dinger
Operators, Standard and Nonstandard (Dubna 1988), World Scientific,
Singapure (1989), pp 79-100.

\bibitem{ExS2}  P. Exner and P. \v {S}eba, Bound states in curved quantum
waveguides, J. Math. Phys. 30(1989), 2574 - 2580.

\bibitem{ExS4}  P. Exner, P. \v {S}eba, Trapping modes in a curved
electromagnetic waveguide with perfectly conducting walls, Phys. Lett. A144
(1990), 347-350

\bibitem{ExV2}  P. Exner and S. A. Vugalter, Asymptotic estimates for bound
states in quantum waveguides coupled laterally through a narrow window, Ann.
Inst. H. Poincare, Phys. Theor. 65 (1996), 109 - 123.

\bibitem{ExV3}  P. Exner, S.A. Vugalter, On the number of particles that a
curved quantum waveguide can bind, J. Math. Phys. 40 (1999), 4630-4638

\bibitem{ExW}  P. Exner, T. Weidl, Lieb-Thirring inequalities on trapped
modes in quantum wires, in \textit{Proceedings of the XIII International
Congress on Mathematical Physics (London 2000)}; to appear [ mp\_arc 00-336]

\bibitem{FW}  M. Freidlin, A. Wentzel, Diffusion processes on graphs and
averaging principle, Ann. Probab., Vol 21, No 4 (1993), pp 2215-2245.

\bibitem{FW1}  M. Freidlin, Markov Processes and Differential Equations:
Asymptotic Problems, Lectures in Mathematics, ETH Zurich, Birkhauser Verlag,
Basel, 1996.

\bibitem{KS}  V. Kostrykin, R. Schrader, Kirchhoff's rule for quantum waves,
J. Phys. A: Mathematical and General, Vol 32, pp 595-630.

\bibitem{K}  P. Kuchment, Graph models of wave propagation in thin
structures, Waves in Random Media, Vol.12, pp 1-24.

\bibitem{K1}  P. Kuchment, Quantum graphs. I. Some basic structures, Waves
in Random Media, Vol.14, No 1 (2004), pp 107-128.

\bibitem{K2}  P. Kuchment, Quantum graphs. II. Some spectral properties of
quantum and combinatorial graphs, Journal of Physics A: Mathematical and
General, Vol 38, No 22 (2005), pp 4887-4900.

\bibitem{KZ1}  P. Kuchment, H. Zeng, Convergence of spectra of mesoscopic
systems collapsing onto a graph, J. Math. Anal. Appl. 258 (2001), PP 671-700.

\bibitem{KZ2}  P. Kuchment, H. Zeng, Asymptotics of spectra of Neumann
Laplacians in thin domains, in Advances in Differential Equations and
mathematical Physics, Yu. Karpeshina etc (Editors), Contemporary
Mathematics, AMS, 387 (2003), PP 199-213.

\bibitem{MV}  S. Molchanov and B. Vainberg, Transition from a network of
thin fibers to quantum graph: an explicitly solvable model, Contemporary
Mathematics, AMS,

\bibitem{MV1}  S. Molchanov and B. Vainberg, Scattering solutions in
networks of thin fibers: small diameter asymptotics, Comm. Math. Phys.,
accepted.

\bibitem{Pavlov4}  Mikhailova, A.; Pavlov, B.; Popov, I.; Rudakova, T.;
Yafyasov, A. Scattering on a compact domain with few semi-infinite wires
attached: resonance case. Math. Nachr. 235 (2002), 101--128.

\bibitem{Pavlov2}  B. Pavlov, K. Robert, Resonance optical switch:
calculation of resonance eigenvalues. Waves in periodic and random media
(South Hadley, MA, 2002), 141--169, Contemp. Math., 339, Amer. Math. Soc.,
Providence, RI, 2003.

\bibitem{P}  O. Post, Branched quantum wave guides with Dirichlet BC: the
decoupling case, Journal of Physics A: Mathematical and General, Vol 38, No
22 (2005), pp 4917-4932.

\bibitem{P1}  O. Post, Spectral convergence of non-compact
quasi-one-dimensional spaces, Ann. Henri Poincar, 7 (2006), pp 933-973.

\bibitem{RS}  J. Rubinstein, M. Schatzman, Variational problems on multiply
connected thin strips. I. Basic estimates and convergence of the Laplacian
spectrum, Arch. Ration. Mech. Anal., 160 (2001), No 4, pp 293-306.
\end{thebibliography}
\end{document}